\magnification=1200 
\baselineskip=15pt

\def\a{\alpha}      \def\l{\lambda}   \def\L{\Lambda} 
\def\b{\beta}       \def\m{\mu}       \def\d{\delta}     
\def\g{\gamma}      \def\G{\Gamma}    \def\n{\nu}        
\def\ve{\varepsilon} \def\o{\omega}   \def\p{\pi}
\def\t{\tau}        \def\s{\sigma}    \def\vphi{\varphi}
\def\ve{\varepsilon}     \def\k{\kappa}  \def\O{\Omega}
\def\cH{{\cal H}}      \def\cP{{\cal P}}  \def\cO{{\cal O}} 
\def\cL{{\cal L}}      \def\cD{{\cal D}}

\def\fr#1 #2{\hbox{${#1\over #2}$}}        
\def\leaderfill{\leaders\hbox to 1em{\hss.\hss}\hfill}
\def\ver#1{\left\vert\vbox to #1mm{}\right.}

\def\section#1{                            
\vskip.6cm\goodbreak                       
\noindent{\bf \uppercase{#1}}
\nobreak\vskip.4cm\nobreak  }

\def\subsection#1{
  \vskip.4cm\goodbreak
  \noindent{\bf #1} 
  \vskip.3cm\nobreak}

\def\subsub#1{\par\vskip4pt {\bf #1} }

\font\eightrm=cmr8                                
\font\tfont=cmbx12                                

\def\title#1{ \centerline{\tfont{#1}} }
\def\titlef#1{ \vskip.2cm                          

      \centerline{ \tfont{#1}
                   \hskip-5pt ${\phantom{\ver{3}}}^\star$  }
      \vfootnote{$^\star$}{\eightrm Work supported in part 
      by the Serbian Research Foundation, Yugoslavia.} \vskip.5cm }
\def\author#1{ \vskip 1cm \centerline{#1} }
\def\institution#1{ \centerline{\it #1} }

\def\abstract#1{ \vskip2.5cm \noindent{\bf Abstract.}\hskip .4cm  
                  {#1} \vfill\eject } 

  \voffset=-9pt
  \vsize=24truecm 
  \hsize=16truecm 
  \hoffset=.2cm
\null
\def\part{\partial}
\def\sh{\hbox{\rm sh$\,$}}                   
\def\ch{\hbox{\rm ch$\,$}}                   \def\tgh{\hbox{\rm tgh$\,$}}
               
\def\ctg{\hbox{\rm ctg$\,$}}                 \def\tg{\hbox{\rm tg$\,$}}   

\hskip2cm

\title{Asymptotic symmetry and conservation laws}
\titlef{in 2d Poincar\'e gauge theory of gravity}
\vskip.7cm
\author{M. Blagojevi\'c}  
\institution{Institute of Physics, 11001 Belgrade, P. O. Box 57, Yugoslavia} 
\author{M. Vasili\'c and T. Vuka\v sinac}
\institution{Institute of Nuclear Sciences Vin\v ca, Department of
             Theoretical Physics,} 

\institution{11001 Belgrade, P. O. Box 522, Yugoslavia}

\abstract{The structure of the asymptotic symmetry in the Poincar\'e
gauge theory of gravity in 2d is clarified by using the Hamiltonian
formalism.  The improved form of the generator of the asymptotic
symmetry is found for very general asymptotic behaviour of phase space 
variables, and the related conserved quantities are explicitly
constructed.}

\subsection{1. Introduction} 

General relativity is a succsessful  theory of macroscopic gravitational
pheno\-mena, but all attempts to quantize the theory encounter serious
difficulties [1]. 
It seems na\-tural to try to understand the structure of gravity on the
basis of the concept of gauge symmetry, which has been very successful
in describing other fundamental interactions in nature.  Poincar\'e gauge
approach to gravity leads to a linear connection with torsion and the
tetrad field as independent variables [2]. The investigation of the
general action of this type, which is at most quadratic in the
curvature and the torsion, shows that the theory does not contain
ghosts or tachions, but the power--counting  renormalizability is
unfortunately lost [3].

One expects that investigations of two--dimensional theories of
gravity may provide a better understanding of quantum properties of
higher dimensional gravity, as well as a deeper understanding of 
string theory. The presence of torsion in string theory considerably
modifies the theory and changes its dynamical content and quantum
pro\-perties [4]. The investigation of classical solutions of the
two--dimensional gravity with dynamical torsion in the conformal gauge
shows that the system is completely integrable [4]. In the light--cone
gauge, which is more appropriate for the separation of true dynamical
degrees of freedom, general analytic solutions of the classical
equations of motion are found [5]. The origin of the classical
integrability is traced back to a specific local symmetry of the theory
found in the Hamiltonian formalism [6]. 
It is interesting to note that the first order formulation of the
theory coincides with the gauge theory based on the nonlinear extension
of the Poincer\'e algebra [7].  Both classical structure of the theory
and its quantum properties show very interesting features [8,9], which
might be helpfull in our attempts to understand  more realistic
four--dimensional theory.  

The Hamiltonian approach was very usefull in clarifying symmetry
properties of the four--dimensional Poincer\'e gauge theory [10--12].
In the present paper we shall use this method to analyse the local
symmetries of the Poincer\'e gauge theory in $2d$, and study the
structure of the symmetry in the asymptotic region. The correct
definition of the asymptotic generators is important for several
reasons: it can be used to study the related conservation laws, to 
find an explicit form of the conserved quantities, and to study the
stability problem.     

We begin our consideration in section 2 by developing the basic
Hamiltonian formalism, and constructing the generator of the local
symmetries. Then we clarify the important notion of the asymptotic
symmetry, which is determined by the behaviour of phase space variables
in the asymptotic region, and discuss the question of surface terms and
the conservation laws. In sections 3 and 4 we use this formalism to
study the asymptotic structure of the theory when the solutions in the
asymptotic region define the space--time of constant curvature and
vanishing torsion, and the Minkowski space--time, respectively. These
specific examples motivate the central consideration in sections 5 and
6, where we derive general rules for constructing the asymptotic
ge\-nerators and the conserved quantities under rather general asymptotic
conditions. We find that the conserved quantities are expressed in
terms of the constant of motion $Q$, known from earlier investigations
[6]. Section 7 is devoted to concluding remarks. Some technical
detailes are displayed in Appendices.   


\subsection{2. Hamiltonian structure of Poincar\'e gauge theory in 2d} %

\subsub{Classical action.} The basic dynamical variables
of this theory are the diad $b^a{_\m}$ and the connection
$A^{ab}{_\m}$, associated with the translation and Lorentz subgroup of
the Poincar\'e group, respectively. Here, $a,b,...=0,1$ are the local
Lorentz indices, while $\m,\n,...=0,1$ are the coordinate indices.  
The structure of the Poincar\'e group is reflected in the existence
of two kinds of gauge field strengths: the torsion  $T^a{_{\m\n}}$, and
the curvature $R^{ab}{_{\m\n}}$. The geometrical structure of the
theory corresponds to the Riemann--Cartan geometry $U_2$.
The most general action of the Poincar\'e gauge theory in 2d, which is
at most quadratic in gauge field strengths, has the form [4]: 
$$
I=\int d^2x b\left( {1\over 16\a}R^{ab}{_{\m\n}}R_{ab}{^{\m\n}}
        -{1\over 8\b}T^a{_{\m\n}}T_a{^{\m\n}}-\g\right) \, ,  \eqno(2.1)
$$
where $b=\det(b^a{_\m})$, and $\a,\b$, $\g$ are constants. The term
linear in curvature is dynamically trivial, as it represents a
topological invariant. In 2d the Lorentz connection can be parametrized
as $A^{ab}{_\m}=\ve^{ab}A_\m$, so that  
$$\eqalign{
&R^{ab}{_{\m\n}}=\ve^{ab}F_{\m\n}\, ,\qquad 
                 F_{\m\n}\equiv\part_\m A_\n -\part_\n A_\m \, ,\cr 
&T^a{_{\m\n}}=\nabla_\m b^a{_\n}-\nabla_\n b^a{_\m} \, ,}
$$
and $\nabla_\m b^a{_\n}=\part_\m b^a{_\n} +\ve^a{_c}A_\m b^c{_\n}$ is the
covariant derivative of the diad field. The action (2.1) is invariant
under the local Poincar\'e transformations with parameters
$\o^{ab}=\ve^{ab}\o$ and $\xi^\l$:
$$\eqalign{
&\d_0 b^a{_\m}=\o\ve^a{_c}b^c{_\m}-\xi^\l{_{,\m}}b^a{_\l} 
                                 -\xi^\l\part_\l b^a{_\m} \, ,\cr
&\d_0A_\m = -\part_\m\o -\xi^\l{_{,\m}}A_\l -\xi^\l\part_\l A_\m \, .}
                                                              \eqno(2.2)
$$

The action (2.1) leads to the following equations of motion [4]:  
$$\eqalign{
&{1\over 2\a}D_\m F^{\m\n}+{1\over 2\b}\ve^{ac}T_{ac}{^\n}=0 \, ,\cr
&{1\over 2\a}F_{ac}F^{\m c}
 +{1\over 2\b}\left( T_{cae}T^{c\m e} +D_\n T_a{^{\n\m}} \right)
 -h_a{^\m}\left({1\over 8\a}F_{bc}F^{bc} 
                 +{1\over 8\b}T_{bce}T^{bce}+\g\right)=0 \, .}
$$
Here, $D_\m$ denotes the covariant derivative acting on both local
Lorentz and coordinate indices. 

It is interesting to note that these equations have a solution
describing the space of constant curvature and zero torsion, 
$F=$ const., $T_{abc}=0$. Indeed, in this case the first equation is
automatically satisfied, while the second one yields
$(1/4\a)F_{bc}F^{bc}-2\g=0$. In particular, $F_{bc}=0$ for $\g=0$. 

\subsub{Constraints and the Hamiltonian.}
The Hamiltonian analysis of the theory will help us to understand its
symmetry properties and clarify the meaning of the conservation laws
[13].   

Let us denote the momenta conjugate to the basic Lagrangian variables
by $\pi_a{^\m}$, $\pi^\m$. Since the curvature and the torsion do not
depend on the velocities $\dot b^a{_0}$, $\dot A_0$, one easily finds
the primary constraints,
$$
\p_a{^0}\approx 0 \, ,\qquad \p^0\approx 0 \, .               \eqno(2.3)
$$
The canonical Hamiltonian density has the form
$$
\cH_c=b^a{_0}G_a-A_0G \, ,                                    \eqno(2.4a)    
$$
where
$$\eqalign{
&G_a=-\ve_{ab}b^b{_1}\, E -(\p_a{^1})'-\ve_{ab}A_1\p^{b1} \, ,\cr
&G = \ve^{ab}b_{b1}\p_a{^1} + (\p^1)' \, ,\qquad
     E\equiv \b \p^{e1}\p_e{^1}-\a\p^1\p^1+\g \, ,   }        \eqno(2.4b)
$$
and prime means the space derivative.
The general Hamiltonian dynamics is described by the total Hamiltonian:
$$
H_T=\int dx^1 (\cH_c+u^a\p_a{^0} +u\p^0 ) \, .                \eqno(2.5)
$$

We shall now demand that all constraints be conserved during the 
time evolution of the system governed by the total Hamiltonian. The
consistency of the primary constraints yields the relations  
$$
\{\p_a{^0},H_T\}=-G_a \, ,\qquad\{\p^0,H_T\}=G \, , 
$$
and we conclude that $G$ and $G_a$ are secondary constraints. Their
algebra is of the form
$$\eqalign{
&\{G_a,G_b\}= -2\ve_{ab}(\b\p^{c1}G_c +\a\p^1 G)\d \, ,\cr
&\{G_a,G\}=\ve_a{^b}G_b\d \, ,\qquad     \{G,G\}=0 \, ,}      \eqno(2.6)
$$
so that further consistency conditions are automatically satisfied.

All the constraints of the system are of the first class. Observe the
parallel between the algebra (2.6) and the related structure in $d=4$ [12].  

\subsub{The gauge generators.}
The Hamiltonian theory described above possesses, by construction, the
local Poincer\'e symmetry. The general method for constructing the
generators of local symmetries has been developed by Castellani [11].
The gauge generators of the local Poincer\'e symmetry have the form 
$$
\G=\int dx^1[\dot\ve(t)\G^{(1)}+\ve(t)\G^{(0)}] \, ,
$$
where $\ve(t)$ are arbitrary parameters, phase space functions 
$\G^{(0)},\G^{(1)}$ satisfy the conditions  
$$ \eqalign{
\G^{(1)}&=C_{PFC} \, ,\cr
\G^{(0)}+\{ \G^{(1)},H_T\}&=C_{PFC}\, ,  \cr
\{\G^{(0)},H_T\}&=C_{PFC} \, ,}                                  
$$
and $C_{PFC}$ means primary first class (PFC) constraint [11]. Here,
the equality sign denotes an equality up to constants and squares (or
higher powers) of constraints. 

It is clear that the construction of gauge generators is based on the
Poisson bracket algebra of constraints. Starting with a suitable
combination of primary first class constraints,
one obtains the following form of the gauge generator:
$$
\G=\int dx^1[\dot\xi{^\m}\G_\m^{(1)} + \xi^\m\cP_\m
             +\dot\o\G^{(1)}+\o S ] \, ,                      \eqno(2.7a)   
$$
where
$$\eqalign{
&\G_\m^{(1)}=b^a{_\m}\p_a{^0}+A_\m\p^0\, ,\qquad \G^{(1)}=\p^0 \, ,\cr
&\cP_0=\cH_T\, ,\qquad
       \cP_1=b^a{_1}G_a-A_1G+(b^a{_0})'\p_a{^0}+A_0'\p^0 \, ,\cr
&S=-G+\ve^{ab}b_{a0}\p_b{^0} \, .}                            \eqno(2.7b)
$$
It is easy to check that the action of the generator ($2.7a$) on the
basic dynamical variables gives a good description of the local
Poincer\'e symmetry. This form of the generator will be the starting
point for our study of the asymptotic structure of the theory and the
related conservation laws. 

In order to simplify the notation we shall use $(x^0,x^1)=(\t,\s)$.

\subsub{Asymptotic symmetry.} The physical content of the notion
of symmetry is determined not only by the symmetry of the action,
but also by the symmetry of the boundary conditions.  

To clarify this statement let us consider a set of solutions of the
field equations which have the following behaviour at large distances: 
$$
b^a{_\m}=\tilde b^a{_\m} + \cO(\tilde b^a{_\m}) \, ,\qquad 
             A_\m=\tilde A_\m +\cO(\tilde A_\m) \, .       \eqno(2.8)
$$ 
This notation means that $\cO(\tilde b^a{_\m})/\tilde b^a{_\m}$ tends
to zero as as $\s\to\infty$, and similarly for $A_\m$.
The quantities $\tilde b^a{_\m}$ and $\tilde A_\m$ are asymptotic
values of $b^a{_\m}$ and $A_\m$ which define an Einstein--Cartan space
$\tilde U$. Let us, further, assume that the space $\tilde U$ has some
isometries described by the Killing vectors $K^\m_i$ ($i\le 3$ in
$d=2$). The form--invariance of the metric under the local translation
with $\tilde\xi^\m=c^iK^\m_i$ ($c^i=$ const.) does not imply the
form--invariance of the diad field $\tilde b^a{_\m}$; instead, the diad
may undergo an additional Lorentz rotation defined by a parameter
$-\tilde\o$, which is linear in $c^i$. Thus, there exists a combination
of the isometry transformation and a Lorentz rotation by $\tilde\o$,
such that $\tilde b^a{_\m}$ is form--invariant,   
$$
\tilde\d_0 \tilde b^a{_\m}=0\, ,\qquad 
                            \tilde\d_0\equiv\d_0(\tilde\o,\tilde\xi^\m)\, .
$$
If the same transformation leaves $\tilde A_\m$ invariant, it
represents the symmetry of the space  $\tilde U$, or the asymptotic
symmetry of a given set of solutions.  

The choice of the asymptotic behaviour for dynamical variables defines
the asymptotic structure of spacetime $\tilde U$, which usually
represents the ground state or the vacuum of the theory. The symmetry
of the action  spontaneously breaks down to the symmetry of $\tilde U$,
which is responsible for the existence of conservation laws.

The existence of isometries of the space $\tilde U$ does not guarantee
the existence of asymptotic symmetries. If these symmetries exist, they
can be obtained from the corresponding local Poincer\'e transformations
by the replacement
$$
\xi^\m(x) \to \tilde\xi^\m=c^i K^\m_i \, ,\qquad \o(x) \to \tilde\o \, .
                                                              \eqno(2.9a)
$$
The generator of these transformations can be obtained from the gauge
generator (2.7) in the same manner. It has the general form
$$
\G_{as}= c^i T_i \, ,                                          \eqno(2.9b)
$$
where the explicit form of $T_i$ depends on the structure of $\tilde U$. 

The symmetry transformation of a dynamical variable in the asymptotic
region is defined by its Poisson bracket with the generator $\G_{as}$.
Therefore, in order to give a precise meaning to the notion of
asymptotic symmetry it is necessary to examine the existence of 
functional derivatives of $\G_{as}$. A functional 
$G[\vphi,\pi]=\int d\s g(\vphi,\partial_1\vphi,\pi,\partial_1\pi)$
is said to have well defined functional derivatives if its variation
can be written in the form
$$
\d G=\int d\s\left[ A(\s)\d\vphi(\s) + B(\s)\d\pi(\s)\right] \, ,
$$
where $\d\vphi_{,1}$ and $\d\pi_{,1}$  are absent. In general this is
not the case with the generator $\G_{as}$,  so that its form should be
improved by adding a nontrivial surface term:
$$
\G_{as}\to \tilde\G=\G_{as}+\O \, .                            \eqno(2.10)
$$
The possibility of constructing the improved generator $\tilde\G$
depends on the structure of the phase space in which the generator
$\G_{as}$ acts.      

We now wish to see how the asymptotic symmetry implies the existence of
certain conserved quantities. Although Castelani's method has been
originally designed to study local symmetries, it can be also used to
obtain useful information about global symmetries. One can easily prove
that a phase space functional $G[\vphi,\pi;\t]$ is the generator of
global symmetries if and only if  the following relations hold true:
$$\eqalign{
&\{ G, H_T\} + \partial G/\partial\t = C_{PFC} \, ,\cr
&\{ G,\phi_s\} \approx 0 \, , }                               \eqno(2.11a) 
$$
where $\phi_s$ are all constraints in the theory, and both $G$ and
$H_T$ are corrected by surface terms, if necessary. 
As we mentioned before, the equality in the first equation is an
equality up to terms which act trivially as symmetry generators, such
as constants, squares of constraints and surface terms, so that its
left--hand side need not vanish. Note that we restrict ourselves to
global symmetries obtained from the corresponding local ones by some
process of parameter fixing, so that $G$ and $H_T$ are first class
constraints, and their Poisson bracket vanishes weakly.
Consequently, the first equation, representing the Hamiltonian form of
the conservation law, implies a weak equality 
$$
dG/d\t \equiv \{ G, H_T\}+\partial G/\partial\t \approx 
     \partial G/\partial\t \approx \partial\O/\partial\t \, , \eqno(2.11b)
$$
where $\O$ is a possible surface term, so that $G$ is conserved if the
time derivative of $\O$ vanishes. By calculating $d\tilde\G/d\t$ one
can check on the existence of the conserved quantities in the
asymptotic region.   

The conservation law for the generator $\tilde\G$ without surface terms
is rather trivial: such generator is given as an integral of
constraints, so that its on--shell value always remains zero. 
{\it Nontrivial conservation law can be obtained only in the presence
of nonvanishing surface terms\/}. In that case the on--shell value of
the conserved quantity is equal to the value of the related surface
term: $\,\tilde\G \approx \O$. 

\subsection{3. Solutions with constant curvature and
               zero torsion at large distances} 

We shall first study the asymptotic symmetry when the solutions of the 
gravitational field equations in the asymptotic region represent the
space of constant curvature and zero torsion, $\tilde V$.    

The symmetries of the Riemann space
$\tilde V$ can be obtained from the corresponding local Poincer\'e
transformations by the replacement of parameters $\o(x)$ and $\xi(x)$
as in Eq.(A.9): 
$$
\xi^\m(x) \to c^i K^\m_i \, ,\qquad \o(x) \to -\xi^0{_{,1}} \, ,
$$
where $c^i$ are constants and $K^\m_i$ are the Killing vectors of
$\tilde V$, defined in (A.7). The generator of these transformations can
be obtained from the gauge generator (2.7) in the same manner. An
explicit calculation yields  
$$
\G_{as}= c^0T_0+c^1T_1+c^2T_2 \, ,                             \eqno(3.1) 
$$
where
$$\eqalign{
&T_0=\int d\s \cP_1 \, ,\cr
&T_1=\cos\t\int d\s \big[\cos\s\G_0^{(1)} +\sin\s (\cP_1+\G^{(1)})\bigr]
     +\sin\t\int d\s \bigl[\cos\s\cP_0 -\sin\s (\G_1^{(1)}-S)\bigr] \, ,\cr
&T_2=\cos\t\int d\s \bigl[\sin\s\G_0^{(1)} -\cos\s (\cP_1+\G^{(1)})\bigr]
     +\sin\t\int d\s \bigl[\sin\s\cP_0 +\cos\s (\G_1^{(1)}-S)\bigr] \, .}
$$

In order to examine the existence of well defined functional
derivatives of the ge\-nerator $\G_{as}$ one should first define the phase
space in which the generator acts. 

The structure of the space $\tilde V$ is described
in Appendix A. Starting from Eqs.(A.8), which describe the form of the
diad field and the connection in a suitable coordinate system, we are
led to consider solutions of the theory (2.1) with the following
asymptotic behaviour:   
$$
b^a{_\m}= {r\over\sin\t}\,\d^a_\m +\cO_\a\, ,\qquad 
                        A_\m=\ctg\t\,\d^1_\m +\cO_\b\, ,   \eqno(3.2)
$$
where $\cO_\a$ denotes a term that decreases like $\vert \s\vert^{-\a}$ 
or faster for large $\s$, $\,\a,\b>0$, and $r=(-4\a\g)^{-1/4}$.

We shall demand that all expressions that vanish on shell have an
arbitrarily fast asymptotic decrease, as no solutions of the equations 
of motion are thereby lost. Thus, the asymptotic behaviour of the
momentum variables will be determined by requiring 
$\pi-\partial\cL/\partial\dot\vphi\sim\hat\cO$, where $\hat\cO$ denotes
a term that decreases sufficiently fast, e. g. like $\cO_3$.
From the definition of momenta and the accepted asymptotic behaviour
(3.2) of the Lagrangian variables one finds:
$$\eqalign{
&\pi_a{^0}=\hat\cO\, ,\qquad \pi_a{^1}=\cO_\a + \cO_\b \, ,\cr 
&\pi^0=\hat\cO\, , \qquad\pi^1={1\over 2\a r^2} +\cO_\a +\cO_\b \, . } 
                                                              \eqno(3.3) 
$$
In a similar manner one can determine the asymptotic behaviour of the
Hamiltonian multipliers $u^a$ and $u$.

After having determined the asymptotic behaviour of all phase space
variables, we are now ready to check on the existence of well defined 
functional derivatives of the generator (3.1). Let us first check on
the differentiability  of $T_0$. By varying  $\cP_1$ one finds 
$$\eqalign{
\d\cP_1&=b^a{_1}\d G_a-A_1\d G +\d b^a{_{0,1}}\pi_a{^0} 
                                          +\d A_{0,1}\pi^0+R \cr
       &=b^a{_1}\d G_a-A_1\d G +(\d b^a{_0}\pi_a{^0} 
                                          +\d A_0\pi^0)_{,1}+R\, . }
$$
Here, terms that contain unwanted variations are explicitly displeyed,
while the remaining, regular terms are denoted by $R$. 
Taking into account the relations $\d G_a=-\d\pi_a{^1}{_{,1}}+R$,  
$\d G=\d\pi^1{_{,1}}+R$, and the asymptotic conditions (3.3), the above
relation becomes 
$$
\d\cP_1=-(b^a{_1}\d\pi_a{^1}+A_1\d\pi^1 +\hat\cO)_{,1}+R \, .
$$
After that the integration over $\s$ leads to
$$
\d T_0=-(b^a{_1}\d\pi_a{^1}+A_1\d\pi^1)
            \vert^{\s\to +\infty}_{\s\to -\infty} +R =R \, .  \eqno(3.4a)
$$
The surface term vanishes since the asymptotic values of all dynamical
variables are fixed constants, so that their variations tend to zero.
Therefore, the generator $T_0$ has well defined functional derivatives.

In a similar way we find
$$
\d T_1 =R \, ,\qquad \d T_2=R \, .                            \eqno(3.4b)  
$$
Compared to $T_0$, the expressions for $T_1$ and $T_2$ contain
additional $\s$--dependent terms, but these are of the form $\sin\s$ or
$\cos\s$, i.e. they are bounded, and the structure of surface terms
remains unchanged. 

Vanishing of surface terms means that $T_0$, $T_1$ and $T_2$ are
well defined generators in the asymptotic region. The structure of the
asymptotic symmetry is described by the corresponding Poisson bracket
algebra. Up to PFC terms one finds:
$$
\{T_0,T_1\}=-T_2\, ,\qquad\{T_0,T_2\}=T_1\, ,\qquad\{T_1,T_2\}=T_0 \, .
                                                              \eqno(3.5)  
$$
Not surpraisingly we obtain the $SL(2,R)$ algebra --- the symmetry
algebra of constant curvature Riemann spaces.  

This symmetry implies, as usual, the existence of certain conserved
quantities. An explicit calculation with the help of Eq.(2.11)
shows that the generators $T_0$, $T_1$ and $T_2$ are conserved 
quantities. These conservation laws are, however, rather trivial,
as all surface terms vanish.

The above analysis of surface terms shows some features which are
important for a deeper understanding of the structure of these terms.
By a slight generalization of the previous discussion one can reach the
following conclusion:  
\vskip.2cm
\item{} if~~$a)$ the parameters $\o(x)$, $\xi^\m(x)$ are {\it bounded}
                 as $\s\to\pm\infty$,~~and 
\item{} ~~~~$b)$ the variations of all dynamical variabes
                 {\it vanish}~when~$\s\to\pm\infty\, ,$ \hfill (3.6) 
\item{} than {\it all surface terms vanish\/}. \par
\vskip.2cm
\noindent In the case of asymptotically flat space we shall relax
the condition $a)$ by considering parameters which are linear functions
of $\s$, and see the consequences on the structure of surface
terms. After that we shall study more general case in which none of the
conditions $a)$, $b)$ is fulfilled.

\subsection{4. Asymptotically flat space}

We assume that the asymptotic structure of space--time is described by
the global Poincer\'e symmetry. Global Poincer\'e transformations can be
obtained from the corresponding expressions for local transformations
by the following replacement of parameters:
$$ \eqalign{
&\o(x)\to -\o \, ,\cr
&\xi^\m (x)\to -\o^\m{_\n}x^\n -\ve^\n \, ,}                  \eqno(4.1)
$$
where $\o^{\m\n}=\ve^{\m\n}\o$ and $\ve^\nu$ are constants.
After that, the generators of global transformations take the form
$$
\G_{as}= \o T-\ve^\n T_\n \, ,                                \eqno(4.2a)
$$
where
$$ \eqalign{
&T_{\mu}=\int d\s \cP_\mu ,\cr
&T=\int d\s(\t\cP_1 +\s\cP_0 -S + b^a{_1}\p_a{^0}+A_1\p^0 )\, .}\eqno(4.2b)
$$

We assume that the asymptotic structure of space--time is determined by
the following behaviour of diads and connection:
$$
b^a{_\m}= \d^a_\m +\cO_1 \, ,\qquad A_\m= \cO_1  \, .   \eqno(4.3a)
$$
The first condition describes the rule by which the metric approaches
the Minkowskian form, and the second one ensures the absolute
paralelism in the asymptotic region.  

From the definition of momenta and the relations $(4.3a)$ we find 
$$\eqalign{
&\p_a{^0}=\hat\cO \, ,\qquad \p_a{^1}=\cO_1, \, \cr
&\p^0=\hat\cO\, ,    \qquad ~\p^1=\cO_1 \, .\cr   }      \eqno(4.3b)
$$

Since the vacuum values of all dynamical variables are constant and
the parameters $\ve^\m$ multiplying $T_\m$ are also constants, it is
clear from (3.6) that 
$$
\d T_\m =R \, .                                               \eqno(4.4) 
$$
An explicit calculation shows the paralell with Eq.(3.4$a$).

In the variation of $T$ the only nontrivial contribution comes from
$\s\cP_0$. By observing the relation 
$\d\cP_0= -(b^a{_0}\d\pi_a{^1}+A_0\d\pi^1)_{,1}+R$, and taking into
account the asymptotic behaviour displayed in (4.3) one finds
$$\eqalign{
&\d T=\int d\s [\s\d\cP_0] + R =  -\d\O +R \, , \cr
&\O\equiv (\s\pi_0{^1})\vert^{\s\to+\infty}_{\s\to-\infty}\,\, .}
                                                              \eqno(4.5a)
$$
This result indicates that there might be asymptotically flat solutions
for which the surface term $\O$ is nonvanishing. However, an
inspection of the constraint $G\approx 0$ easily shows that $\pi_0{^1}$
must decrease as $\cO_2$, so that 
$$
\O=0 \, .                                                      \eqno(4.5b)
$$

The asymptotic conditions (4.3) are not the most general conditions
corresponding to the flat space at large distances. It is, therefore,
natural to try to find out whether one can change these conditions in a
way consistent with the Minkowskian structure in the asymptotic region,
and obtain nonvanishing surface terms. The general discussion in the
next section will show that this is not possible.  

\subsection{5. General asymptotic structure} 

We have seen that the conditions $a)$ and $b)$ in (3.6) are of special
importance in considerations of the structure of surface terms, as
they imply vanishing of these terms. The investigation of the
asymptotically flat solutions, which are characterized by linearly
rising parameters when $\s\to\pm\infty$ and, therefore, violate the
condition $a)$, also leads to vanishing surface terms.  This situation
provides a rationale for trying to understand what happens if the
variations of dynamical variables in the asymptotic region do not
vanish, i. e. when the condition $b)$ is also violated. 

\subsection{\it 5.1 The light--cone gauge} 

The generators of the local symmetry (2.7) are constructed so as to act
on both physical and unphysical dynamical variables. In order to
simplify further discussion we shall fix the gauge and eliminate
unphysical variables from the theory. It is clear that this procedure
does not change physical properties of the theory. 

It is convenient to choose the local Lorentz frame in the form of the
light--cone basis:    
$$
a=(+,-)\, ,\qquad \eta_{+-}=\eta_{-+}=1\, ,\qquad \ve^{+-}=-1\, , 
                                     \qquad u\cdot v=u_+v_-+u_-v_+ \, .
$$

The existence of the first--class constraints (2.3) enables us to
fix the values of $(b^a{_0},A_0)$ by imposing the light--cone gauge
conditions:
$$
b^+{_0}=1 \, ,\qquad b^-{_0}=0 \, ,\qquad A_0=0 \, .          \eqno(5.1)
$$
The consistency of the gauge conditions implies 
$$
u^a=0 \, ,\qquad u=0 \, .
$$
After introducing the preliminary Dirac brackets corresponding to the
gauge conditions (5.1) and the constraints (2.3), we can easily see
that the variables $(b^a{_0},A_0,\pi_a{^0},\pi^0)$ become ignorable:
they can be replaced by their values given in Eqs.(2.3) and (5.1),
while the Dirac brackets for the remaining variables coincide with the
Poisson brackets.   

In the light--cone gauge the constraints and the Hamiltonian take the
simpler form: 
$$\eqalign{
&G_+=-b^-{_1}\,E -\p_+{^1}A_1 -\pi_+{^1}{_{,1}} \, , \cr
&G_-= b^+{_1}\,E +\p_-{^1}A_1 -\pi_-{^1}{_{,1}} \, , \cr
&G  = b^-{_1}\p_-{^1} - b^+{_1}\p_+{^1} + \p^1{_{,1}} \, ,\cr               
& H_T=\int d\s G_+ \, , \qquad
      E\equiv 2\b\p_+{^1}\p_-{^1}-\a\p^1\p^1+\g \, , }        \eqno(5.2)
$$
Observe also that now $b=-\ve_{ab}b^a{_0}b^b{_1}=-b^-{_1}$, so that
the condition of the nondege\-neracy of the metric becomes $b^-{_1}\ne 0$.

The equations of motion for the remaining set of variables take the form:
$$
\eqalign{ &\dot\pi_+{^1}=0\, , \cr  
          &\dot b^+{_1}=-2\b b^-{_1}\,\pi_-{^1}-A_1 \, ,} 
   \qquad \eqalign{ &\dot\pi^1=\pi_+{^1} \, ,\cr
                    &\dot A_1=2\a b^-{_1}\,\pi^1 \, ,} 
       \qquad \eqalign{ &\dot b^-{_1}=-2\b b^-{_1}\,\pi_+{^1} \, ,\cr
                        &\dot\pi_-{^1}=E \, .}                          
$$
The first three equations are easily solved:
$$
\pi_+{^1}=A(\s)\, ,\qquad \pi^1=A(\s)\t +B(\s) \, ,
   \qquad b^-{_1}=C(\s)e^{-2\b A(\s)\t} \, ,
$$
where $A(\s)$, $B(\s)$ and $C(\s)$ are three arbitrary functions. The
remaining variables $\pi_-{^1}$, $b^+{_1}$ and $A_1$ can be determined
from the constraints. 

\subsub{} From the form of the local Poincer\'e
symmetry one concludes that the light--cone gauge is preserved if 
the parameters $\dot\xi^1$, $\dot\o$ and $\dot\xi^0+\o$ vanish. After
that, there remains the {\it residual symmetry} defined by parameters 
$$
\xi^1=\ve^1(\s)\, ,\qquad\o=\ve(\s)\, ,
            \qquad \xi^0=-\ve(\s)\t+\ve^0(\s)\, .            \eqno(5.3)
$$
The corresponding generator is obtained by replacing these parameters
into the general expression (2.7) for $\G$:
$$
\G = \int d\s\left[\ve^0\bar\cP_0+\ve^1\bar\cP_1
                                -\ve \bar S \right]\, ,      \eqno(5.4a)
$$
where  
$$\eqalign{
&\bar\cP_0 = G_+\, ,\cr
&\bar\cP_1 = b^+{_1}G_+ + b^-{_1} G_- -A_1 G 
 = -(b^+{_1}\pi_+{^1}{_{,1}} + b^-{_1}\pi_-{^1}{_{,1}}
                                            + A_1\pi^1{_{,1}})\, ,\cr
&\bar S=\t\bar\cP_0 +G \, .}                                  \eqno(5.4b)
$$

To examine the differentiability of $\G$ we calculate $\d\G$: 
$$\eqalign{
\d\G =& \int d\s\left[ -\ve^0 \d\pi_+{^1}{_{,1}} 
 -\ve^1 (b^+{_1}\d\pi_+{^1}+b^-{_1}\d\pi_-{^1}+ A_1\d\pi^1)_{,1}\right.\cr
&\hskip1cm\left. -\ve (\d\pi^1-\t\d\pi_+{^1})_{,1} \right] +R   
           = L\vert^{+\infty}_{-\infty}+R \, ,  }             \eqno(5.5a)
$$ 
where the surface term $L$ is given by the expression
$$
L\equiv  -\ve^0 \d\pi_+{^1} 
         - \ve^1(b^+{_1}\d\pi_+{^1}+b^-{_1}\d\pi_-{^1}+A_1\d\pi^1) 
         - \ve (\d\pi^1-\t\d\pi_+{^1}) \, .                   \eqno(5.5b) 
$$
Explicit form of $L$ depends on asymptotic values of both dynamical
variables and multipliers. If $L$ can be brought into the form 
$L=-\d\O$, the improved generator in the asymptotic region will have
the form (2.10). 

\subsection{\it 5.2 Asymptotic symmetry} 

General relation between boundary conditions and the structure of the
asymptotic symmetry will be given in two steps.

\subsub{A.} Let us first consider the asymptotic conditions
$$
\eqalign{ & b^-{_1}= a^- + \cO(a^-) \, ,\cr
          & b^+{_1}= a^+ + \cO(a^+) \, ,\cr
          & A_1    = a   + \cO(a) \, ,  }   \qquad
       \eqalign{ & \pi_-{^1}= c_- + \cO(c_-) \, ,\cr
                 & \pi_+{^1}= c_+ + \cO(c_+) \, ,\cr
                 & \pi^1    = c   + \cO(c) \, ,  }            \eqno(5.6)
$$
where $\l_\a=(a^\mp, a, c_\mp, c)$ are real parameters independent of
$\s$ and $\t$ which may take different values as $\s\to \pm\infty$,
and $\cO(\l)$ are terms that vanish when $\s\to\pm\infty$. We assume that
the phase space contains a collection of solutions of the type (5.6),
where the real parameters $\l_\a$ are not fixed constants, but belong
to certain subsets of real line.  This, in particular, means that these
parameters can be varied: {\it the variations of dynamical variables in
the asymptotic region do not vanish\/}. 

The restriction to fixed parameters $\l_\a$ would convert the phase
space into a configuration which is usually called vacuum. Here we are
working with a phase space consisting of a collection of different vacua.
This situation is possible as the notion of gauge symmetry enables us
to consider a symmetry transformation that maps a solution of the type
(5.6) into another solution of the same type, but with different $\l_\a$'s.

It is natural to assume that the assymptotic behaviour is restricted by
the demand that all the constraints  decrease sufficiently fast:
$$
G_+\, ,G_-\, , G =\hat\cO\, .                                 \eqno(5.7a)
$$
In particular, the values of the parameters $\l_\a$ are restricted by
demanding the consistency with the constraints.  Thus, one can take the
values of $a^-$, $c_+$ and $c$ arbitrarily,  while $a^+$, $a$ and $c_-$
are restricted to satisfy the constraints $G$, $G_+$ and $G_-$: 
$$\eqalign{
&a^+c_+=a^- c_- \, ,\cr
&a c_+ =-a^- (2\b c_+ c_- -\a c^2 +\g) \, ,}                  \eqno(5.7b)  
$$
Let us note that the condition (5.7) also restricts the form of
different factors $\cO$ appearing in (5.6).

We will restrict our discussion to the case 
$$
b^-{_1}\ne 0 \, ,                                             \eqno(5.8) 
$$
so as to ensure the nondegeneracy of the metric in the whole
space--time.  

By a simple inspection of the expression (5.5) for the surface term one
concludes that the Hamiltonian and the generators 
$\int d\s\ve^0\bar\cP_0$, $\int d\s\ve\bar S$ can easily be made well
defined (finite and differentiable) by adding a suitable surface term,
if  $\ve^0$ and $\ve$ satisfy the conditions 
$$
\ve^0,\ve = \hbox{\rm const.} + \cO \, .                      \eqno(5.9)
$$ 

In order to investigate the generator $\int d\s\ve^1\bar\cP_1$, we
note that  
$$
L=-\d\left[\ve^0\pi_+{^1} + \ve(\pi^1 - \t\pi_+{^1})\right]
  -\ve^1 ( b^+{_1}\d\pi_+{^1} + b^-{_1}\d\pi_-{^1} + A_1\d\pi^1 ) \, .
                                                             \eqno(5.10)
$$

The first part has the form of the variation of a surface term, while
the second one requires further discussion.
Let us first consider the simple case when $c_+\ne 0$. Then  by using
the constraints (5.7) one finds (Appendix B)
$$\eqalign{
&b^+{_1}\d\pi_+{^1} + b^-{_1}\d\pi_-{^1} + A_1\d\pi^1
   = -{\a\over 8\b^3}\,{a^-\over c_+}e^{2\b c}\d\tilde Q \, ,\cr
&\tilde Q \equiv e^{-2\b c}\left[ 1+(1+2\b c)^2 
        -{4\b^2\over\a}(2\b c_+ c_- +\g )\right]\, ,} 
$$ 
and the expression for $L$ becomes   
$$
L=-\d \left[ \ve^0\pi_+{^1} +\ve(\pi^1-\t\pi_+{^1}) \right]
   +{\a\over 8\b^3}\,\ve^1\,{a^-\over c_+}\,e^{2\b c}\d\tilde Q \, .
$$
The consistency of the above discussion requires $\ve^1=$ const. $+\cO$.

The above result for $L$ does not have the form of a total variation,
so that the form of the generator $\G_{as}$ cannot be improved. The
problem can be solved by redefining the parameter $\ve^1$ as follows:   
$$
\ve^1={\pi_+{^1}\over b^-{_1}}e^{-2\b\pi^1}\eta \, ,
                   \qquad \eta=\hbox{\rm const.} + \cO \, .   \eqno(5.11)
$$
Then, without using the assumption $c_+\ne 0$, the calculation
analogous to the one given above leads to 
$$
L=-\d\left[\ve^0\pi_+{^1} +\ve(\pi^1-\t\pi_+{^1})
    - {\a\over 8\b^3}\,\eta\,\tilde Q \right] \, .            
$$
It is now easy to see from Eq.(5.5$a$) that the variation of the
generator can be written in the form
$$\eqalign{
&\d\G_{as} = -\d\O + R \, ,\cr
&\O\equiv \left[\ve^0\pi_+{^1} +\ve(\pi^1-\t\pi_+{^1})
  - {\a\over 8\b^3}\,\eta\,\tilde Q \right]^{+\infty}_{-\infty}\, ,}
                                                              \eqno(5.12) 
$$
where the surface term $\O$ is defined on the boundary of the
one--dimensional space. Now we can redefine the generator $\G_{as}$ as
in Eq.(2.10), $\tilde\G=\G_{as}+\O$, so that $\tilde\G$ has well defined
functional derivatives. The assumed asymptotic behaviour ensures
finitness of $\O$. The generator $\tilde\G$ does not vanish on--shell as
$\G_{as}$ does --- it takes on the value $\O$. The related conservation
law will be discussed in the next section. 

The above result for the surface term is obtained in the phase space
defined by (5.6). The residual symmetry of the theory is defined by
the parameters that are bounded at large distances. It is clear
from the form of the general surface term (5.5) that the asymptotic
symmetry, defined by the behaviour of parameters at large distanceses,
is closely related to the form of the asymptotic conditions. More
general asymptotic symmetries can be obtained by considering more
general asymptotic conditions.

\subsub{B.} The previous considerations can be easily generalized
so as to include solutions with more general asymptotic behaviour. 

$(i)$ Let us first generalize the asymptotics (5.6) by assuming that
for large $\s$ the dynamical variables behave as
$$
\eqalign{ & b^-{_1}= M^- + \cO(M^-) \, ,\cr
          & b^+{_1}= M^+ + \cO(M^+) \, ,\cr
          & A_1    = M   + \cO(M) \, ,  }   \qquad
       \eqalign{ & \pi_-{^1}= N_- + \cO(N_-) \, ,\cr
                 & \pi_+{^1}= N_+ + \cO(N_+) \, ,\cr
                 & \pi^1    = N   + \cO(N) \, ,   }        \eqno(5.13)
$$ 
where $(M^\mp,M,N_\mp,N)$ are certain functions of $\s$, not
necesserily bounded in $\s$.    

$(ii)$ We shall demand that all constraints  decrease sufficiently
fast, as in Eq.(5.7$a$). 

$(iii)$ The condition of the nondegeneracy of the metric (5.8) will be
also retained.

$(iv)$ Leaving the question of the finitness of various integrals for
the end, one can redefine the parameter $\ve^1$ as in Eq.(5.11). Then,
starting from the expression (5.10) for $L$ and making  use of the
constraints (following the calculation outlined in Appendix B) one
obtaines the result
$$\eqalign{
&\d\G_{as} = -\d\O + R \, ,\cr
&\O\equiv \left[\ve^0\pi_+{^1} +\ve(\pi^1-\t\pi_+{^1})
  - {\a\over 8\b^3}\,\eta\, Q \right]^{+\infty}_{-\infty}\, ,}
                                                              \eqno(5.14) 
$$
where $Q$ is defined in $(B.3)$.
This leads to the correctly defined generator $\tilde\G=\G_{as}+\O$ if
$\eta\sim$ const. The result (5.14) is obtained provided the following
condition is satisfied:    
$$
\eta e^{-2\b\pi^1}(b^-{_1})^{-1}\bigl(
\pi^1{_{,1}}\d\pi_+{^1} -\pi_+{^1}{_{,1}}\d\pi^1\bigr)=\cO\, .   
$$
This condition is ensured if we demand 
$$\eqalignno{
& \pi_+{^1}\, ,\; \pi^1 \; \sim \; \hbox{\rm const.} \, ,&(5.15a)\cr
&b^-{_1}\quad\hbox{\rm does~not~decrease~faster~than}\quad\cO_1 \, .& 
                                                                 (5.15b)} 
$$
The geometrical meaning of these conditions is discussed in Appendix C.

The last term in $\O$ gives a nonzero contribution if $\eta\ne 0$ and
the values of $\eta$ when $\s\to\pm\infty$ are different.

\subsection{\it 5.3 Asymptotically flat space --- general consideration} 

The general structure of asymptotic symmetries obtained so far gives
us a clue to understand, on more general grounds, the properties 
of space--time which at large distances has the Minkowskian
structure: 
$$\eqalign{
&b^+{_1}=0   \, , \qquad  b^-{_1}=1  \, ,\qquad A_\m=0 \, ,\cr
&\pi_+{^1}=0 \, , \qquad \pi_-{^1}=0 \, ,\qquad \pi^1=0 \, .} 
                                                              \eqno(5.16)
$$

The symmetries of the Minkowski vacuum (5.16) are obtained by the
following restriction on the parameters $\ve^\m(\s),\ve(\s)$:
$$\eqalign{
&\ve(\s)=\ve=\hbox{\rm const.}\, ,
         \qquad\ve^0(\s)=\ve^0=\hbox{\rm const.} \, ,\cr
&\ve^1(\s)=\ve\s+\ve^1 \, ,\qquad \ve^1=\hbox{\rm const.} }   \eqno(5.17a) 
$$
They define the global Poincer\'e transformations, with the generator
$$\eqalign{
&\G_{as}=\ve^0 T_0 +\ve^1 T_1 -\ve T \, ,\cr
& T_\m=\int d\s\bar\cP_\m \, , \qquad
        T=\int d\s (\t\bar\cP_0 -\s\bar\cP_1 + G) \, . }      \eqno(5.17b)
$$

We choose the following asymptotic behaviour:
$$\eqalign{
&b^+{_1}=\cO \, ,\qquad b^-{_1}= 1+\cO \, ,\qquad A_1=\cO \, ,\cr
&\pi_+{^1}=\cO\, ,\qquad\pi_-{^1}=\cO \, ,\hskip36pt\pi^1=\cO\, ,}
                                                              \eqno(5.18)
$$
where $\cO$ stands for the general term of the $\cO_\a$ type, with $\a>0$.

The expression for the surface term $L$ simplifies:
$$
L=-\ve\s(b^+{_1}\d\pi_+{^1}+b^-{_1}\d\pi_-{^1} +A_1\d\pi^1) +\cO
 = -\ve\s\pi^1{_{,1}}\d\ln\pi_+{^1} +\cO  \, ,
$$
where we used $\d Q=\hat\cO$ (in vacuum $Q=2$; since $Q'$ is a
constraint, one concludes that $Q=2+\hat\cO$, therefore $\d Q=\hat\cO$). 
By using the relation 
$$
\s\pi^1{_{,1}}\d\ln\pi_+{^1}=
   \d(\s\pi^1{_{,1}}\ln\pi_+{^1})-\s(\d\pi^1){_{,1}}\ln\pi_+{^1} \, ,
$$ 
we easily see that $L=0$ under the asymptotic conditions (5.18).

\subsection{6. Conserved quantities --- the construction} 

Now that we have the general expression (5.14) for the surface term
we can improve the symmetry generators and make them finite and
differentiable. Thus, we define
$$
\tilde\G=c^0\tilde T_0 + c^1\tilde T_1 -c\tilde T \, ,        \eqno(6.1)
$$
where $c^0$, $c^1$ and $c$ are constants, and the improved generators
are given by
$$\eqalign{
&\tilde T_0=\int d\s(\ve^0\bar\cP_0) +\O_0\, ,\hskip75pt
   \O_0\equiv (\ve^0\pi_+{^1})\vert^{+\infty}_{-\infty} \, ,\cr
&\tilde T_1=\int d\s\left( \eta{\pi_+{^1}\over b^-{_1}}\,
                           e^{-2\b\pi^1}\bar\cP_1\right)+\O_1\, ,  \qquad
   \O_1\equiv -{\a\over 8\b^3}(\eta Q)\vert^{+\infty}_{-\infty} \, ,\cr
&\tilde T =\int d\s(\ve\bar S) + \O \, ,  \hskip93pt
   \O\equiv [\ve(\pi_+{^1}\t-\pi^1)]\vert^{+\infty}_{-\infty} \, .}
                                                              \eqno(6.2)
$$
Here, $\ve^0(\s)$, $\ve^1(\s)$ and $\eta(\s)$ are arbitrarily given
functions of $\s$, the asymptotic behaviour of which is chosen to
ensure 
\item{$a)$} the finitness of the generators, and 
\item{$b)$} the invariance of the asymptotic conditions  (5.13).

For a given set of parameters $\ve^0(\s)$, $\ve^1(\s)$ and $\eta(\s)$,
the functionals (6.2) are the global symmetry generators and,
consequently, satisfy the equations (2.11), where the Hamiltonian is
also corrected by a surface term:
$$
\tilde H_T =\int d\s \bar\cP_0 +(\pi_+{^1})\vert^{+\infty}_{-\infty}\, ,
                                                              \eqno(6.3) 
$$
The Poisson bracket $\{\tilde\G,\tilde H_T\}$, being basically the
commutator of two first class constraints, vanishes on--shell. It
follows then that $d\tilde\G/d\t\approx \partial \tilde\G/\partial\t$, 
which means that only those generators whose surface terms have an
explicit time dependence may not be conserved. 

For the generators $\tilde T_0$ and $\tilde T_1$ one easily finds the
relations 
$$
{d\tilde T_0\over d\t}\approx {\partial\O_0\over\partial\t}=0\, ,\qquad
{d\tilde T_1\over d\t}\approx {\partial\O_1\over\partial\t} =0\, ,
                                                              \eqno(6.4a)
$$
showing that the surface terms $\O_0$ and $\O_1$ represent the values of
$\tilde T_0$ and $\tilde T_1$ as the conserved charges.
Any particular choice of $\ve^0$ and $\ve^1$ defines the related
expressions for $\O_0$ and $\O_1$, respectively. In particular, if
$\ve^0=1$ then  $\tilde T_0=\tilde H_T$, and  
$\O_0$ represents the conserved value of the Hamiltonian. Taking into
account the existence of parameters $\ve^0$ and $\ve^1$ with various
asymptotic behaviour, we find that the following quantities are also
conserved: 
$$
\O_0^\pm\equiv\pi_+{^1}(\s\to\pm\infty)\, ,\qquad
    \O_1^\pm\equiv Q(\s\to\pm\infty)\equiv Q^\pm\, .          \eqno(6.4b)
$$
Note also that from $Q'\approx 0$ it follows that $Q(\s)$ is a constant
of motion with the value $Q^+=Q^-$ for all $\s$'s.

As concerns the generator $\tilde T$, we find  
$$
{d\tilde T\over d\t}\approx {\partial\O\over\partial\t}= 
      (\ve \pi^1_+)\vert^{+\infty}_{-\infty}\, ,               \eqno(6.5)
$$
so that $\tilde T\approx\O$ will be conserved if $\O_0^\pm=0$.  

Close inspection of the equations of motion justifies the results of
the above symmetry considerations. 

\subsection{7. Concluding remarks} 

We presented here an analysis of the relation between the asymptotic
symmetries and the conservation laws in the Poincer\'e gauge theory in
$2d$. The generators of the asymptotic symmetries are constructed from
the corresponding gauge generators. By demanding that the asymptotic
generators have well defined functional derivatives we obtained their
improved form, containing certain surface terms. These surface terms
represent the values of the conserved quantities, whose physical
meaning depend on the nature of the asymptotic symmetry. The analysis
is carried out for very general asymptotic conditions of phase space
variables. 

It can be generalized to other gravitational theories in $2d$, such as
the dilaton gravity, and used to study the stability problem.

\subsection{Appendix A: The Riemann space of constant curvature in $2d$}

1. The Riemann space of constant curvature is maximally symmetric
space $\tilde V$ [14]. By using the fact that maximally symmetric spaces
are essentially unique, one can find the metric of such space in $2d$ by
considering a two--dimensional hypersphere  embedded in a
three--dimensional flat space: 
$$
ds^2=\eta_{\m\n}dx^\m dx^\n +\k^{-1}dz^2\, , \qquad 
     \k\eta_{\m\n}x^\m x^\n +z^2=1 \, ,\qquad \k=\hbox{\rm const.}
$$
After eliminating $z$, the metric of $\tilde V$ takes the form  
$$
ds^2=g_{\m\n}dx^\m dx^\n \, ,\qquad 
   g_{\m\n}\equiv\eta_{\m\n}+{\k x_\m x_\n\over 1-\k x^2}\, , \eqno(A.1)
$$
where $x^2=\eta_{\m\n}x^\m x^\n$. The isometries of $\tilde V$ are
defined by the following Killing vectors: 
$$
\xi^\m_{(a)}=\d^\m_a\sqrt{1-\k x^2} \quad(a=0,1)\, ,
                \quad\qquad \xi^\m_{(2)}=\ve^\m{_\n}x^\n \, . \eqno(A.2)
$$

2. For $\k\le 0$  it is convenient to introduce new coordinates,
$$\eqalign{
&x^0= r\sh\t  \, , \quad\qquad r^2=1/\vert\k\vert \, ,\cr
&x^1=r\ch\t\cos\s \, , \cr
&z= \ch\t\sin\s \, ,}
$$
where $\t\in(-\infty,+\infty)$, $\s\in(0,2\pi)$.
The new form of the metric is given by
$$
ds^2=r^2(d\t^2-\ch^2\t d\s^2 ) \, .                           \eqno(A.3)
$$
It is interesting to note that here we can enlarge the domain for $\s$
by assumming $\s\in(-\infty,+\infty)$.
The Killing vectors in new coordinates are given as:
$$
\xi^\m_{(0)} = {1\over r}\pmatrix{\sin\s      \cr
                                  \tgh\t\cos\s \cr} \, ,\qquad
\xi^\m_{(1)} = -{1\over r}\pmatrix{ 0 \cr
                                    1 \cr} \, ,\qquad
\xi^\m_{(2)} = \pmatrix{-\cos\s     \cr
                        \tgh\t\sin\s \cr} \, .                \eqno(A.4)
$$
They satisfy the Killing equation in the whole region
$\t\in(-\infty,+\infty)$,  $\s\in(-\infty,+\infty)$.

3. Let us now introduce another coordinate transformation, 
$$
\t=\ln\tg(\t'/2) \, ,\qquad \s=\s'\, , \quad\qquad \t'\in(0,\pi)\, ,
                     \quad \s'\in(-\infty,+\infty) \, , 
$$
which transforms the metric into conformally flat form:
$$
g'_{\m\n}(\t',\s')={r^2\over\sin^2\t'}\eta_{\m\n} \, .        \eqno(A.5)
$$
The Killing vectors are given as
$$
\xi'^\m_{(0)} = {1\over r}\pmatrix{\sin\t'\sin\s' \cr
                             -\cos\t'\cos\s' \cr} \, ,\quad
\xi'^\m_{(1)} = -{1\over r}\pmatrix{ 0 \cr
                              1 \cr} \, ,\quad
\xi'^\m_{(2)} = -\pmatrix{ \sin\t'\cos\s' \cr
                           \cos\t'\sin\s' \cr} \, .           \eqno(A.6) 
$$
After changing the basis
$$
K_0=-r\xi'_{(1)} \, ,\qquad K_1=-\xi'_{(2)}\, ,\qquad K_2=r\xi'_{(0)}\, ,
                                                              \eqno(A.7) 
$$
one easily finds
$$
[K_0,K_1]=-K_2 \, ,\qquad [K_0,K_2]=K_1\, ,\qquad [K_1,K_2]=K_0 \, .
$$
The symmetry of the space is characterized by the $SL(2,R)$ algebra, as
expected. 

4. We now discuss the form of the diad and the connection by using the 
conformally flat form of the metric, and dropping primes for simplicity. 
The fields $b^a{_\m}$ and $A_\m$ will be defined by demanding
$T^a{_{\m\n}}=0$ and $R^{ab}{_{\m\n}}=$ const. 

From $T^a{_{\m\n}}=0$ one finds the relation
$A^\m=\ve^{ab}h_b{^\n}h_a{^\m}{_{,\n}}$. After that one concludes from
(A.5) that the diad and the connection can be choosen in the form:  
$$
b^a{_\m}={r\over\sin\t}\,\d^a_\m \, ,\qquad 
                     A_\m=\ctg\t\,\d_\m^1 \, .                \eqno(A.8) 
$$
Explicit check leads to $F=h_a{^\m}h_b{^\n}R^{ab}{_{\m\n}}=-2/r^2$.

The isometry transformatios $\xi^\m =c^iK^\m_i$, where $c^i$ are
constants, do not change the form of the metric (A.5).
What happens with the fields $b^a{_\m}$ and $A_\m$? Introducing the
notation $\d_0^K=\d_0(\o=0,\xi^\m=a^iK^\m_i)$ one finds
$$
\d_0^K b^a{_\m}=-\sin\t(a^1\sin\s-a^2\cos\s)\ve^a{_c}b^c{_\m}
                   = \xi^0{_{,1}}\ve^a{_c}b^c{_\m} \, .
$$
Thus, the diad field is not invariant under the isometries. However,
the change $\d_0^K b^a{_\m}$ has the form of a local Lorentz
transformation. Combining the isometry transformation with
the local Lorentz transformation defined by $\o=-\xi^0{_{,1}}$ one
obtaines  
$$
\d_0b^a{_\m}=0 \, .\qquad \d_0 A_\m = 0 \, .
$$
Therefore, the Poincer\'e transformations defined by the global parameters 
$$
\xi^\m =a^iK^\m_i\, ,\qquad \o=-\xi^0{_{,1}}\, ,              \eqno(A.9)
$$ 
leave the diad and the connection (A.8) of the space $\tilde V$ invariant.

\subsection{Appendix B: Surface term and constant of motion} 

In this Appendix we shall derive the expression (5.12) for the surface
term, and show that it is given in terms of a constant of motion. We
start from the relation  
$$\eqalign{
L_1&\equiv b^+{_1}\d\pi_+{^1} + b^-{_1}\d\pi_-{^1} + A_1\d\pi^1
 =a^+\d c_+ + a^-\d c_- +a\d c +\cO  \cr
&={a^-\over c_+}c_-\d c_+ +a^- \d c_-
  -{a^-\over c_+}(2\b c_  + c_- -\a c^2 +\g)\d c +\cO  }      \eqno(B.1)
$$
obtained on the basis of the assymptotic behaviour (5.6) and the
constraints (5.7). The expression on the right hand side can be written
in the form   
$$\eqalign{
&{a^-\over c_+}\left[\d(c_-c_+)-(2\b c_  + c_- -\a c^2 +\g)\d c\right]
 =-{\a\over 8\b^3}{a^-\over c_+}\left[ \d\L -2\b\L\d c\right] \, ,\cr
&\L\equiv 1+(1+2\b c)^2 -{4\b^2\over\a}(2\b c_+ c_- +\g) \, .}
$$
It is now easy to see that 
$$
L_1=-{\a\over 8\b^3}{a^-\over c_+}e^{2\b c}\d\tilde Q +\cO \, ,\qquad
 \tilde Q \equiv e^{-2\b c}\L \, .                            \eqno(B.2)
$$
whereafter the result (5.12) follows directly. 

It is interesting to note that $\tilde Q$ represents the asymptotic
value of 
$$
Q\equiv e^{-2\b\pi^1}\left[ 1+(1+2\b\pi^1)^2 
       -{4\b^2\over\a}(\g +2\b\pi_+{^1}\pi_-{^1})\right]\, .  \eqno(B.3)
$$
The quantity $Q$ is related to an important feature of the dynamics of
the theory, which consists in the existence of a general constant of
motion in the phase space. To see that we observe the relation:
$$
\pi^{a1}G_a + EG = {\a\over 8\b^3}e^{2\b\pi^1} Q' \, ,        \eqno(B.4)
$$
Since $Q'\approx 0$, it follows that $Q$ is a function of
time only,  $Q=Q(\t)$. It is easy to check that $\{Q,G_a\}=\{Q,G\}=0$,
therefore  $\dot Q=\{Q,H_T\}=0$, i. e. $Q$ is a constant of motion: 
$$
Q=Q_0=\hbox{\rm const.}                                       \eqno(B.5)  
$$

The constraints $G$, $G_+$ and $Q'$ are equivalent to $G$, $G_+$ and
$G_-$. 

\subsection{Appendix C: The geometrical meaning of the conditions (5.15)}

Let us note, first, that the value of the variable $\pi^1$
coincides with the curvature of the space--time. The equation of motion
$\dot\pi^1=\pi_+{^1}$ then tells us that the asymptotic behaviour
$\pi^1,\pi_+{^1}\sim$ const. is necessary if we want the space--time to
have asymptotically finite curvature:
\vskip.2cm
\item{}~~~~$(5.15a)$~~$\Longleftrightarrow$~~
            asymptotically finite curvature. \hfill           $(C.1)$
\vskip.2cm
\noindent At the same time, the condition $\pi_+{^1}\sim$ ~const. ensures
finiteness and differentiability of the Hamiltonian.

To understand the asymptotic behaviour of $b^-{_1}$ we shall consider
the quantity
$$
\cD\equiv \int d\s\left({b^-{_1}\over \pi_+{^1}}\right)e^{2\b\pi^1} \, ,
$$
which is proportional to the length of the lines along which the
space--time curvature is constant (see the third reference in [9]). The
asymptotic condition (5.15$b$) then implies that $\cD$ is infinite: 
\vskip.2cm
\item{}~~~~$(5.15b)$~~$\Longleftrightarrow$~~infinite
                           lines of constant curvature. \hfill $(C.2)$
\vskip.2cm
\noindent The integral $\cD$ is a natural measure of the linear size of
space--time, since it is obviously a gauge invariant quantity.

\subsection{References}                

\item{1.} G. t'Hooft and M. Veltman, Ann. Inst. H. Poincer\'e, {\bf 20}
  (1974) 245; M. H. Goroff and Sagnotti, Nucl. Phys. {\bf B266} (1986)
  709. 
\item{2.} T. W. B. Kibble, J. Math. Phys. {\bf 2} (1961) 212.
  F. W. Hehl, P. von der Heide, G. D. Kerlick and J. M. Nester,
  Rev. Mod. Phys. {\bf 48} (1976) 393.
\item{3.} E. Sezgin and P. van Nieuwenhuizen, Phys. Rev. {\bf D21}
  (1980) 3269.
\item{4.} I. V. Volovich, Phys. Lett. {\bf B175} (1986) 413;
  Two--dimensional gravity with dynamical torsion and 
  strings, preprint CERN--TH 4771/87 (1987); 
M. O. Katanaev and I. V. Volovich, Ann. of Phys. (NY) {\bf 197} (1990) 1;  
M. O. Katanaev, Theor. Math. Phys. {\bf 80} (1989) 838; 
  J. Math. Phys. {\bf 31} (1990) 882; {\bf 32} (1991) 2483.
\item{5.} W. Kummer and D. J. Schwarz, Phys. Rev. {\bf D45} (1992) 3628; 
  Nucl. Phys. {\bf B382} (1992) 171. 
\item{6.} H. Grosse, W. Kummer, P. Pre\v snajder and D. J. Schwarz, 
  J. Math. Phys. {\bf 33} (1992) 3892;
\item{7.} N. Ikeda and Ken--Iti Izawa, Prog. Theor. Phys. {\bf 89}
  (1993) 223; {\bf 89} (1993) 1077.
\item{8.} K. G. Akdeniz, A. Kizilers\" u and E. Rizao\^ glu, Phys.
  Lett. {\bf B215} (1988) 81;
K. G. Akdeniz, \" O. F. Dayi and A. Kizilers\" u, Mod.
  Phys. Lett {\bf A7} (1992) 1757;
\item{9.} T. Strobl, Int. J. Mod. Phys. {\bf A8} (1993) 1383;
P. Schaller and  T. Strobl, Phys. Lett.{\bf B337} (1994) 266;
                       Class. Quant. Grav. {\bf 11} (1994) 331.
W. Kummer and D. J. Schwarz, Renormalization of $R^2$ Gravity
  with Dynamical Torsion in $d=2$, preprint TUW--91--09r (1991);
F. Heider and W. Kummer, preprint TUW--92--15 (1992);
\item{10.} M. Blagojevi\'c and I. Nikoli\'c, Phys. Rev.  {\bf D28}
  (1983) 2455; I. Nikoli\'c, Phys. Rev. {\bf D30} (1984) 2508.
M. Blagojevi\'c and M. Vasili\'c, Phys. Rev {\bf D36} (1987) 1679; 
M. Blagojevi\'c, I. Nikoli\'c and M. Vasili\'c, Nuovo Cimm. {\bf B101}
  (1988) 439;  
\item{11.} L. Castellani, Ann. Phys. NY {\bf 143} (1982) 357.
\item{12.} M. Blagojevi\'c and M. Vasili\'c, Class. Quant. Grav. 
  {\bf 5} (1988) 1241.
\item{13.} P. A. M. Dirac, {\it Lectures on Quantum Mechanics}
  (Yeshiva University, New York, 1964). 
A. Hanson, T. Regge i C. Teitelboim, {\it Constrained Hamiltonian
  Systems} (Aca\-demia Nazionale dei Lincei, Rome, 1976).
K. Sundermeyer, {\it Constrained Dynamics} (Springer, Berlin, 1982). 
\item{14.} S. Weinberg, {\it Gravitation and Cosmology\/}
   (John Wiley and Sons, New York, 1972). 

\bye